# Standardized Descriptive Index for Measuring Psychometric Indicator Deviation and Uncertainty

**Mark Dominique Dalipe Muñoz**

Iloilo Science and Technology University – Main Campus

**Abstract**

**Introduction**: The use of descriptive statistics in pilot testing procedures requires objective, standard diagnostic protocols that are feasible for small sample sizes. While current psychometric practices report item-level statistics, they often report these raw descriptives separately rather than consolidating both mean and standard deviation into a single diagnostic tool to directly measure item quality.

**Methods**: By leveraging the analytical properties of Cohen's $d$, this article repurposes its use in scale development as a standardized item deviation index. This measures the extent of an item's raw deviation relative to its scale midpoint while accounting for its own uncertainty.

**Results**: Analytical properties such as boundedness, scale invariance, and bias are explored to further understand how the index values behave.

**Conclusion**: This will aid future efforts to establish empirical thresholds that characterize redundancy among formative indicators and consistency among reflective indicators.

*Keywords*: Psychometrics, Survey questionnaire, Cohen's $d$, Descriptive Statistics, Estimation, Item Analysis

# 1. Standardized Descriptive Index for Measuring Psychometric Indicator Deviation and Uncertainty

## 1.1. Background of the Study

Achieving universal interpretability on psychometric measurements is a challenging constraint when operationalizing abstract concepts. Unlike well-established physical quantities such as temperature, weight, and height, scales for latent constructs are prone to subjective anomalies and errors that compromise construct reliability and validity (Price, 2017; Nunnally, 1967). How can one measure something whose truth is often relative to an individual or a specific context? Because of their inherent abstractness, the goal is to establish standardized measures to ensure reliable and valid results that mirror the true characteristics of the attributes as closely as possible (Nunnally, 1967; Cooper, 2023). For years, scholars have developed diagnostic protocols to assess the quality of scale indicators (Swan et al., 2023; Downing & Haladyna, 2006; Hays & Carew, 2021), yet their application remains inconsistent and misconceived across field sciences (Swanson, 2014; Wijsen et al., 2021). One instance is the historical dominance of Classical Test Theory (CTT) in assessing reflective constructs, which has left formative constructs as relatively less understood and often addressed only within advanced statistical frameworks like Structural Equation Modelling (SEM) (Churchill, 1979; Jarvis et al., 2003; Diamantopoulos & Winklhofer, 2001). These formative constructs are also frequently encountered in marketing, information theory, and econometrics, which further amplify the perceived "academic silos" involving these fields (Diamantopoulos & Winklhofer, 2001; Freeze & Raschke, 2007; Henseler et al., 2009; Jarvis et al. 2003).

Proposed formative frameworks currently highlight the use of descriptive statistics as a tool to assess item quality and redundancy among formative indicators. This is to ensure that selected items are functional enough to discriminate responses across participants without introducing instrumental bias due to content or scope within the construct's domain (Muñoz, 2025; Price, 2017). However, these guidelines remain heuristic, and further efforts to establish their objective, standard guidelines are currently explored. If using descriptive statistics is necessary to measure item redundancy, at what point is an item empirically too redundant? Also, pilot testing procedures often have small sample sizes that introduce computational limitations among "gold standard" psychometric tools such as Exploratory Factor Analysis and Variance Inflation Factor ($VIF$) (Muñoz, 2025; Kyriazos, 2018; Marcoulides & Raykov, 2018; Dalawi et al., 2025; O'brien, 2007). This highlights a need for model-free diagnostic tools that are easily accessible and measurable, given that descriptive statistics among samples are one of the statistical measures that are straightforward to compute using pilot sample data alone. While item-level statistics do report descriptive statistics, they are reported as separate raw measures and hence, a diagnostic tool that consolidates diagnostic information regarding the item deviation and variability provides a more efficient approach on determining item quality (Price, 2017; McCowan & McCowan 1999; Ellis, 2017).

Hence, this article proposes a standardized item deviation index that measures the extent to which the measured mean deviates from its theoretical midpoint while accounting for the variability across responses. This ensures the value is comparable across all other items to inform researchers on what indicator to address first, depending on the nature of the psychometric model assumed. While the author intends the proposed tool's application towards formative models to measure item redundancy, its use on reflective models is plausible and yet to be explored, since these psychometric models are viewed as opposites of one another, where the goal shifts from measuring redundancy to consistency. We first proceed with proving its *statistical moments* and *origin* as an unscaled t-statistic that converges to a z-score at larger sample sizes. We then define and explore three of its analytical properties: (1) *Boundedness*: Does a finite domain exist in either the formula itself or its inputs? (2) *Scale Invariance*: Does changing the nature of the scale affect how we interpret the measure? (3) *Bias Adjustment*: Because our index is known to overestimate like Cohen's $d$, what would the corrected index be expressed as? Finally, a *discussion* regarding the implications of the analytical outcomes and results will be presented to aid future directions of the proposed formula.

**1.2. Estimation**

In estimation theory, our goal is to find an approximate value of a particular true value $\theta$ (Doob, 1936). Given a set of observations $x \in X$ and we want to estimate the parameter $\theta \in \Theta$, we define the following function:

$$\hat{\theta}: X \rightarrow \Theta \text{ such that } \hat{\theta}(x) \approx \theta \tag{1}$$

is called an estimator (Gentle, 2020) whereas:

$\theta$ : estimand, true value
$\hat{\theta}$ : estimator, function
$\hat{\theta}(x)$ : estimate, estimated value

Simply put, an estimator is like a particular "rule" or "expression" that takes the actual data as input to produce a corresponding estimate of the true value (Gentle, 2020; Lundberg et al., 2021).

Furthermore, a bias in estimation is the statistical discrepancy between the expected value of the estimator and the estimand defined as:

$$\text{Bias}(\hat{\theta}) = E[\hat{\theta}(x)] - \theta$$

In other words, an estimator is biased if it fails to converge to the true value $E\left[\hat{\theta}(x)\right] \neq \theta$, where it either overestimates (Bias > 0) or underestimates (Bias < 0). Otherwise, an estimator is said to be unbiased $E\left[\hat{\theta}(x)\right] = \theta$ (Andrews et al., 2025; Gentle, 2020; Delacre et al., 2021).

### 1.3. Statistical Moments

Statistical moments are numerical measures that describe the characteristics of the probability distribution of any random variable or dataset. They can be thought of as "blueprints" of the data's "statistical personality". There are three common variants defined depending on the point around which they are calculated.

For a random variable $X$, the $k$-th moment are formally defined as follows (Gentle, 2020):

- $\mu'_k = E[X^k]$: Raw moment describes the distribution characteristics relative to the origin and are both location-dependent (varies by $\mu$ and scale-dependent $\sigma^2$.
- $\mu_k = E[(X - \mu)^k]$: Central Moment describes the distribution characteristics relative to the population mean and are scale-dependent.
- $\lambda_k = \frac{\mu_k}{\sigma_k}$: Standardized Moment are dimensionless units of measurement on the distribution characteristics.

**Table 1**

*Summary of Statistical Moments*

| | $\mu'_k$ | $\mu_k$ | $\lambda_k$ |
|---|---|---|---|
| First Moment ($k = 1$) | $E[X]$ | 0 | 0 |
| Second Moment ($k = 2$) | $E[X^2]$ | $E[(X - \mu)^2]$ | 1 |
| Third Moment ($k = 3$) | $E[X^3]$ | $E[(X - \mu)^3]$ | $\frac{\mu_3}{\sigma_3}$ |

| | $\mu'_k$ | $\mu_k$ | $\lambda_k$ |
|---|---|---|---|
| Fourth Moment ($k = 4$) | $E[X^4]$ | $E[(X-\mu)^4]$ | $\frac{\mu_4}{\sigma_4}$ |

In this article, we focus on first raw moment (mean), second central moment (variance), and third and fourth standardized moments (skewness and kurtosis) as key measures to describe $P(X)$ (see Table 1) because they carry the most straightforward interpretation on their values (Gupta et al., 2019; Demir, 2022).

Some literature and in some statistical softwares such as IBM SPSS, however, prefer $\frac{\mu_4}{\sigma_4} - 3$ instead as the fourth standardized moment (Kim, 2013). This is known as the excess kurtosis and is centered at zero for easier interpretation since negative and positive values are effectively distinguished.

**1.4. Standard Normal Distribution**

Standard Normal Distribution ($Z \sim N(0,1)$), also known as Z-distribution, is a continuous probability distribution defined by $E[Z] = 0$ and $Var[Z] = 1$. For any normally distributed random variable $X$ with mean $\mu$ and standard deviation $\sigma$ (Howell, 2010; Gentle, 2020), we can transform any observation $x \in X$ into a corresponding z-score using the following formula:

$$z = \frac{x - \mu}{\sigma} \tag{2}$$

The resulting Z-scores $z \in Z$ now collectively form the Standard Normal distribution. Its shape and density are mathematically defined by its Probability Density Function (PDF), denoted as $\phi(z)$.

$$\phi(z) = \frac{1}{\sqrt{2\pi}} e^{-\frac{z^2}{2}}$$

The probability curve of the Standard Normal Distribution is therefore generated by plotting all points $(z, \phi(Z))$, where $z$ and $\phi(z)$ are the x- and y-coordinates, respectively. In addition, a Z-statistic can also be calculated directly using the mean of your sample data when the goal is to statistically infer whether the value is significantly different from the standard normal distribution as the theoretical distribution by comparing against the critical values of Z-distribution defined by $\phi(z)$. This informs whether a particular sample assumes a standard normal distribution as the underlying probability distribution and is mathematically expressed as follows:

$$z = \frac{\overline{x}_n - \mu}{\frac{\sigma}{\sqrt{n}}} \qquad \text{where Standard Error } (SE) = \frac{\sigma}{\sqrt{n}} \tag{3}$$

**Figure 1**

*Standard Normal (Z) Distribution (author-generated via ggplot2 R package)*

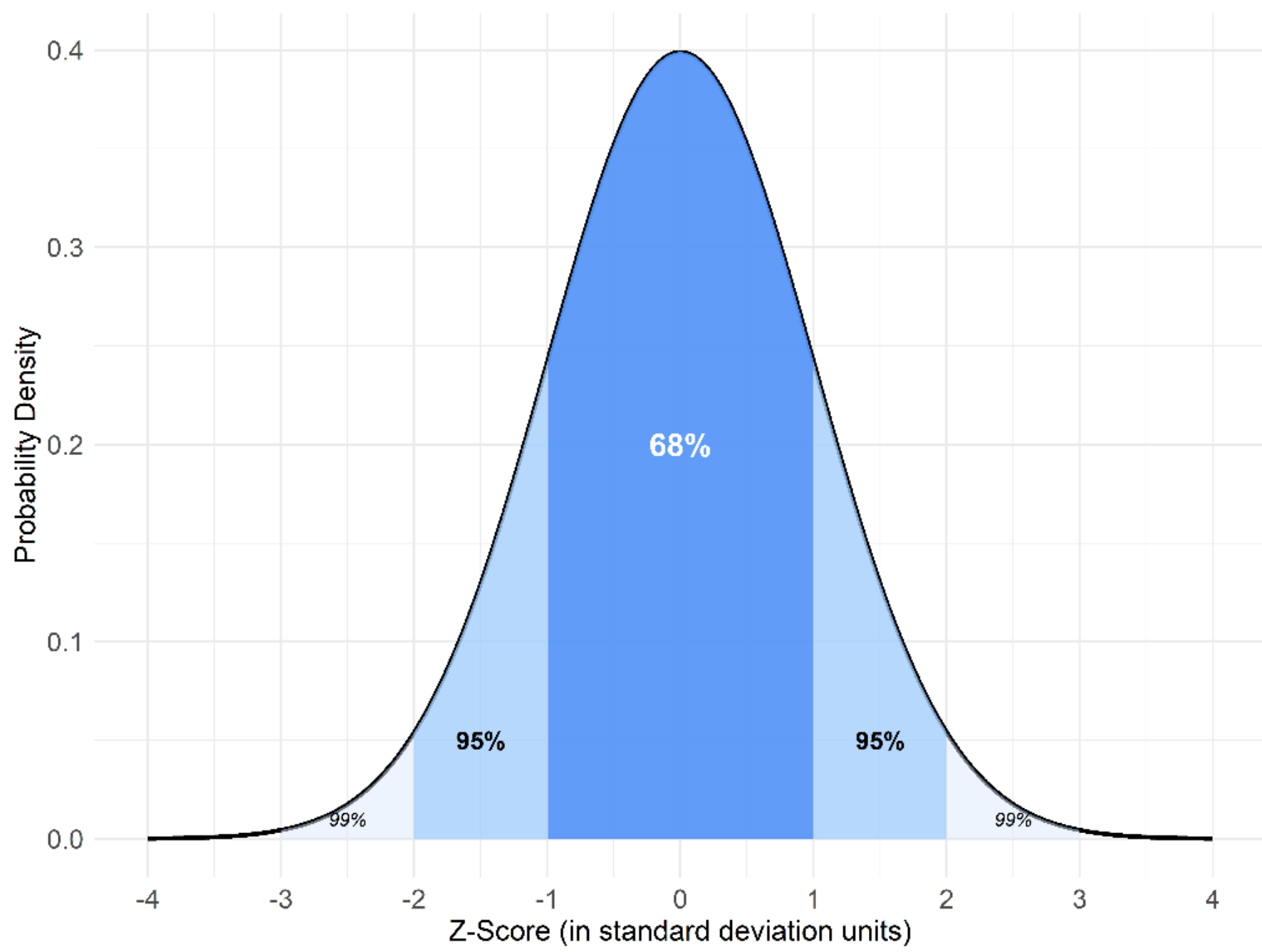

In the figure above, a $Z$-distribution generally looks like a "bell-curve" centered at 0, where $\phi(z)$ represents the black curve. The different color shading on the area under $\phi(z)$ represents the proportion of data that falls within the specified domain in $x$-axis. In other words, we can see that 68% of the scores fall within 1 SD away from the center ($\mu = 0$), 95% of the scores fall within 2 SD away from $\mu$, and approximately 99% of the scores fall within 3 SD away.

## 1.5. Central $t$-Distribution

A Central t-Distribution is a common version of Student's t-distribution for random variables when the null hypothesis is true (see Hypothesis Testing). It is the type of distribution that converges to the Standard Normal Distribution at larger sample sizes (Kirkby et al., 2025).

It is a continuous probability distribution defined by $E[T] = 0$ and $Var[T] = \frac{\nu}{\nu-2}$ where $\nu = n - 1$. For any normally distributed random variable $X$ with mean $\mu$ and standard deviation $\sigma$, we can transform any observation $x \in X$ into a corresponding t-score using the following formula:

$$t = \frac{\overline{x} - \mu}{\frac{s}{\sqrt{n}}} \tag{4}$$

The resulting T-scores $t \in T$ now collectively form the Central t-Distribution. Its shape and density are mathematical defined by its PDF, denoted as $\phi(t)$. Hence, the probability curve of the Central $t$-Distribution is generated by plotting all points $\big(t, \phi(t)\big)$, where $t$ and $\phi(t)$ are the $x$- and $y$-coordinates, respectively.

$$\phi(t) = \frac{\Gamma\left(\frac{\nu + 1}{2}\right)}{\Gamma\left(\frac{\nu}{2}\right)} \cdot \left(\frac{\sigma}{\nu\pi}\right)^{\frac{1}{2}} \cdot \left(1 + \frac{\sigma}{\nu}(t - \nu)^2\right)^{-\frac{\nu+1}{2}}$$

**Figure 2**

*Comparison of Z-distribution and $t$-distribution at selected degrees of freedom (author-generated via ggplot2 R package)*

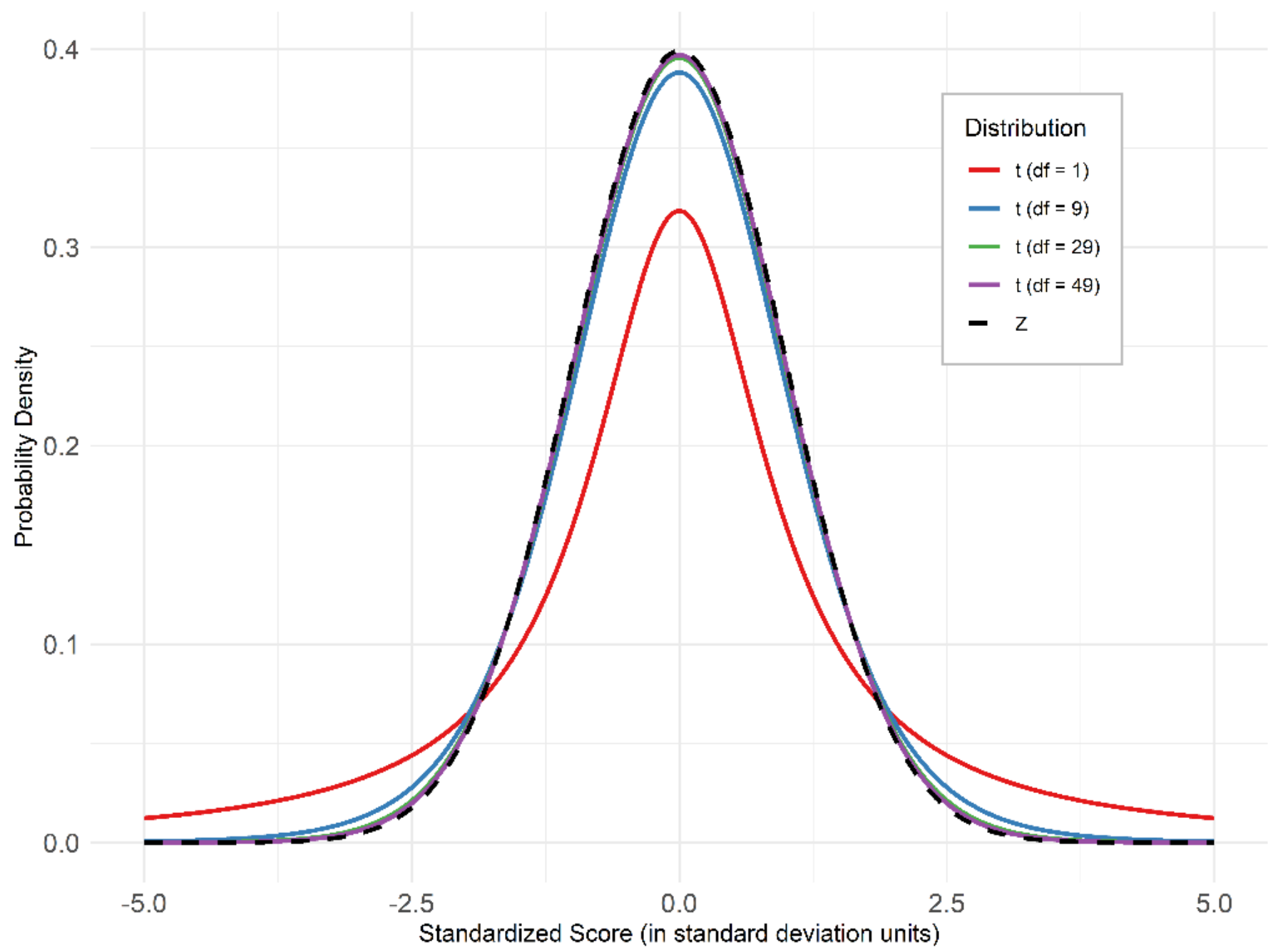

Here, we see $\phi(t)$ curves at different degrees of freedom: 1, 9, 29, 49, which corresponds to having a sample size of 2, 10, 30, and 50, respectively. A higher sample size is observed on $\phi(t)$ to gradually converge ("be more similar") to $\phi(z)$. By default, a $t$-distribution typically have "thicker" tails to be more conservative on extreme uncertainty at small sample sizes.

### 1.6. Null Hypothesis Significance Testing

Null Hypothesis Significance Testing (NHST) is a type of formal hypothesis testing framework for statistical inference that uses data from a sample to test the statistical significance of an assumed hypothesis (Howell, 2010; Pernet, 2017). It involves two competing hypothesis, namely:

- Null Hypothesis ($H_0$): This hypothesis assumes a statement of equality or no effect, difference, or change. It is the default hypothesis that NHST always assumes.
- Alternative Hypothesis ($H_1$ or $H_a$): This hypothesis assumes a statement of difference or change and may either be directional or non-directional. It is the hypothesis that the researchers are seeking to test its significance to conclude.

Hence, the goal of NHST is to determine whether the observed change from the given data has enough statistical significance for $H_0$ to be rejected, implying to support $H_1$. To formally test this property, there are two ways to perform (Howell, 2010): through Confidence Intervals (Neyman-Pearson's approach), where we compare our test statistic to the critical values of the distributon or p-values (Fisher's approach), where we compare our obtained p-value to the predefined alpha ($\alpha$) value instead.

**Figure 3**

*Null Hypothesis Significance Testing (author-generated via ggplot2 R package)*

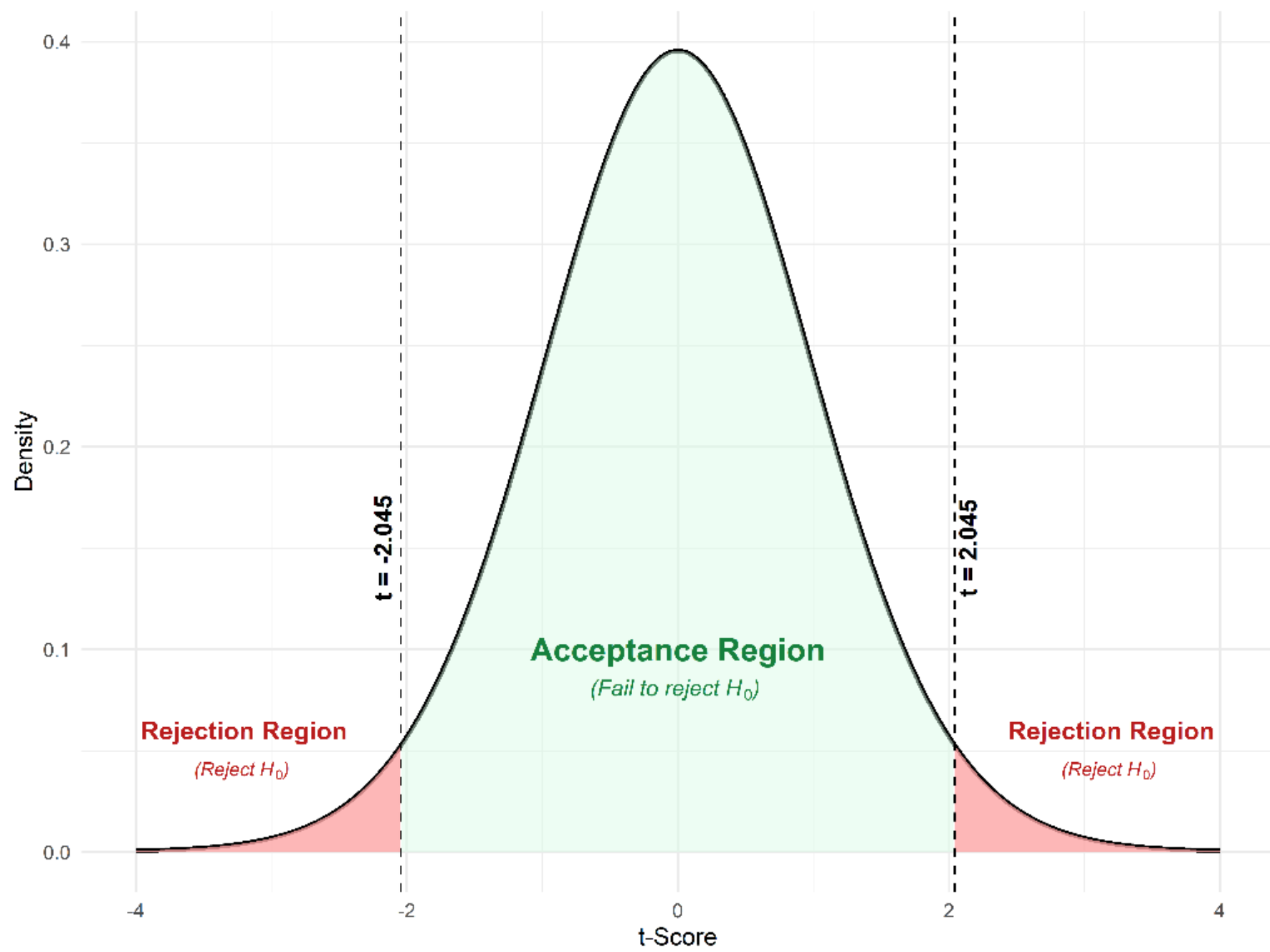

We show an example of a $t$-distribution with 30 observations ($\nu = 29$) to better understand NHST from Neyman-Pearson's approach. At 5% level of significance ($\alpha = 0.05$, the two-tailed critical values $t = \pm 2.046$ establishes a threshold for statistical significance. Given its center to be equal to 0 ($\mu = 0$), a t-score that stays within the acceptance or at the critical threshold means that we fail to reject $H_0$. This is analogous to having a non-significant $p$-value from Fisher's approach. Otherwise, a t-score situated at either side of the rejection regions means we reject $H_0$ and is analogous to a significant $p$-value from Fisher's approach.

### 1.7. Linderberg-Lévy Central Limit Theorem

While there are different versions of Central Limit Theorem (CLT), introductory literature commonly refer to this type of CLT (Inlow, 2010), where the sampling distribution must not only be independent but also identically distributed (i.i.d.).

Let $\{X_n\}$ be a sequence of $n$ i.i.d. random variables from a population with finite moments: $E[X_n] = \mu$ and $Var[X_n] = \sigma^2$.

$$\lim_{n \to \infty} = P(Z_n \leq z) = \Phi(z)$$

where:

$P(Z_n \leq z)$: CDF of normalized sample mean
$\Phi(z)$: CDF of standard normal distribution

In other words, the sample mean $\overline{X}_n$ is assumed to be normally distributed as the sample size ($n$) becomes sufficiently large (Bettini, 2024).

$$\overline{X}_n \sim N\left(\mu, \frac{\sigma^2}{n}\right) \tag{5}$$

**1.8. Slutsky's Theorem**

Let $\{X_n\} = \frac{\overline{x}_n - \mu}{\frac{\sigma}{\sqrt{n}}}$ and $\{Y_n\} = \frac{s_n}{\sigma}$ be two sequences of random variables where $X$ is a random variable and $c$ is a constant. Given that the following conditions are true:

1. Convergence in Distribution: $X_n \overset{d}{\to} X$ by Linderberg-Lévy Central Limit Theorem where $X = Z$.
2. Convergence in Probability: $Y_n \overset{p}{\leftarrow} c$ by Weak Law of Large Numbers (WLLN) where $c = 1$. WLLN ensures that as sample size $n$ increases, $Y_n$ converges to a probability of 1.

Thus, one of the following properties is true for limiting (asymptotic) distributions (Gentle, 2020):

$$\frac{X_n}{Y_n} \overset{d}{\to} \frac{X}{c}$$

$$\left(\frac{\overline{x}_n - \mu}{\frac{\sigma}{\sqrt{n}}} \cdot \frac{1}{\frac{s_n}{\sigma}}\right) \overset{d}{\to} \frac{Z}{1}$$

$$T \xrightarrow{d} Z \tag{6}$$

This proves that t-statistic ($T$) approximates the Z-statistic ($Z$) at larger sample sizes.

### 1.9. Popoviciu's Inequality

Popoviciu (1933) posits that the theoretical maximum bound on the variance is dependent on its total range. Suppose we have $x_i \in X$ observations where $i = 1,2, \dots, n$. Let the arithmetic mean

$$\overline{x} = \frac{1}{n}\sum_{i=1}^{n} x_i,$$

variance

$$s^2 = \frac{1}{n}\sum_{i=1}^{n} (x_i - \overline{x})^2,$$

and range $r$, Then (Sharma et al., 2010),

$$s^2 \leq \frac{r^2}{4} \tag{7}$$

### 1.10. Min-max Normalization

Min-Max Normalization is a common technique to transform a given set of observations $x_i \in X$ into values that strictly fall within the domain $[0,1]$ (Mazziota & Pareto, 2021; Kappal, 2019; Patro & Sahu, 2015)

$$x_{norm} = \frac{x_i - x_{min}}{x_{max} - x_{min}} \tag{8}$$

where:

$x_i$ :raw observation
$x_{min}$ :minimum value within set *X*
$x_{max}$ :minimum value within set *X*

### 1.11. Cohen's $d$ for One-Sample $t$-Test

Cohen's $d$ is a descriptive effect size measure to interpret the magnitude of difference between two means (Goulet-Pelletier & Cousineau, 2018). In one-sample t-test with a degrees of freedom ($\nu$) of $n - 1$ (Howell, 2010), it is defined as comparing the deviation of the sample mean from a selected target value while accounting for the standard deviation to allow comparison across other variables.

$$d = \frac{\overline{x} - \text{Target Value}}{s} \tag{9}$$

where:

$\overline{x}$:sample mean
$s$:sample standard deviation

The target value set depends on the intent of the formula, but its usual convention is the population mean $\mu$.

### 1.12. Hedges's Correction

Cohen's $d$ is known to be a biased estimate since it overestimates the true effect size when sample sizes are small, in which Hedges (1981) proposed a correction factor $J(\nu)$ that adjusts the said bias. The derived value, Hedges' $g$ is therefore a robust estimator of the standardized mean measure.

$$J(\nu) \approx 1 - \frac{3}{4\nu - 1} \tag{10}$$

### 1.13. Shannon's Entropy for Continuous Distributions

Shannon (1948) defined Entropy ($H$) measures the amount of uncertainty of a given information that tells the absolute minimum number of bits needed to store or transmit a message. For continuous values with a standard deviation $\sigma$, it is defined as:

$$H(x) = \log\sqrt{2\pi e}\sigma \tag{11}$$

### 1.14. Continuous Function Theorem for Sequences

The Continuous Function Theorem for Sequences in calculus (Stewart, 2016) bridges the concepts on convergence in sequence and continuity of functions. Let $\{a_n\}$ be a given sequence of values. If $\lim_{n\to\infty} a_n = L$ and the function $f$ is continuous at $L$, then

$$\lim_{n\to\infty} f(a_n) = f(L) \tag{12}$$

## 2. Methods

### 2.1. Derivation of the Index

Given Equation 2, we define True Deviation $\Delta$ for an item $i \in \mathbb{Z}^+$ as a random variable that measures the relative deviation of a raw score to its theoretical midpoint in terms of standard deviation units.

$$\Delta_i = \frac{X_i - M}{\sigma_i}$$

where:

$X_i$:raw score of a random observation for item $i$
$\sigma_i$:true standard deviation of item $i$
$M$:theoretical midpoint of the scale

Since $\Delta$ is related to the Figure 1, we show that $\Delta_i \sim \mathcal{N}(0,1)$.

Let $\mu_i, M \in \mathbb{R}$ and $\sigma_i > 0$ such that $Var[\mu_i] = Var[M] = Var[\sigma_i] = 0$.

To show that the first moment $E[\Delta_i] = 0$:

$$
\begin{aligned}
E[\Delta_i] &= E\left[\frac{X_i - M}{\sigma_i}\right] \\
&= \frac{1}{\sigma_i}(E[X_i] - E[M]) && E\left[\frac{1}{\sigma_i}\right] = \frac{1}{\sigma_i} \\
&= \frac{1}{\sigma_i}(\mu_i - M) && E[X_i] = \mu_i; E[M] = M \\
&= \frac{1}{\sigma_i}(M - M) && \mu_i = M \text{ under } H_0 \\
E[\Delta_i] &= 0
\end{aligned}
$$

To show that the second moment $Var[\Delta_i] = 1$:

$$
\begin{aligned}
Var[\Delta_i] &= Var\left[\frac{X_i - M}{\sigma_i}\right] \\
&= \left(\frac{1}{\sigma_i}\right)^2 Var[X_i] && Var[aX] = a^2\, Var[X] \\
&= \frac{1}{\sigma_i^2} Var[X_i] && Var[X_i - M] = Var[X_i] \\
&= \frac{1}{\sigma_i^2} \cdot \sigma_i^2 && Var[X_i] = \sigma_i^2 \text{ by definition} \\
Var[\Delta_i] &= 1
\end{aligned}
$$

Since $\Delta_i$ assumes a standard normal distribution, Equation 5 allows us to pivot from the distribution of the individual scores $X_i$ to the distribution of their averages ($\overline{\Delta}$).

Hence,

$$\overline{\Delta}_n = E[\Delta_i] = 0$$

$$Var[\overline{\Delta}_n] = \frac{Var[\Delta_i]}{n} = \frac{1}{n}$$

Using Equation 3, we will express it in terms of $\Delta$.

$$
\begin{aligned}
Z &= \frac{\overline{\Delta} - E[\Delta]}{SE} \\
&= \frac{\overline{\Delta} - 0}{SE} && SE = \sqrt{Var[\overline{\Delta}]} \\
&= \overline{\Delta} \cdot \sqrt{n}
\end{aligned}
$$

We then derive the expression of $\overline{\Delta}$.

$$
\begin{aligned}
\overline{\Delta} \quad &= \frac{1}{n}\sum_{k=1}^{n}\rho_k \\
&= \frac{1}{n}\sum_{k=1}^{n}\frac{X_k - M}{\sigma_i} && E[\overline{\Delta}] = E[\Delta_k] = \Delta_k \\
&= \frac{1}{n\sigma_i}\left(\sum_{k=1}^{n}X_k - \sum_{k=1}^{n}M\right) && E[\sigma_i] = \sigma_i \\
&= \frac{1}{\sigma_i}\left(\frac{1}{n}\sum_{k=1}^{n}X_k - \frac{1}{n}\sum_{k=1}^{n}M\right) \\
\overline{\Delta} \quad &= \frac{\overline{x} - M}{\sigma_i} && \frac{\sum_{k=1}^{n}X_k}{n} = \overline{x}; \frac{\sum_{k=1}^{n}M}{n} = M
\end{aligned}
$$

Hence, substituting $\overline{\Delta}$ on $Z$ gives us:

$$Z = \left(\frac{\overline{x} - M}{\sigma_i}\right) \cdot \sqrt{n}$$

Although $Z$ is an ideal formula one can use to calculate the deviation of an item mean to the theoretical midpoint, it cannot be calculated analytically because it requires a standard deviation from a population. In the context of pilot testing, the number of respondents does not involve the population because the motive is to refine and improve the contextual and psychometric properties of the questionnaire before the actual data gathering. Hence, $Z$ remains to be a theoretical concept and thus, we need an alternative statistic that acts as a proxy for the the properties of $Z$.

By Equation 6, we can derive an alternative formula based on Equation 4. Given its raw moments as follows:

$$E(t) = 0 \text{ and } Var(t) = \frac{\nu}{\nu - 2} \qquad \text{where } \nu > 2$$

We can express $t$ as follows:

$$
\begin{aligned}
t &= \frac{\overline{x} - \mu}{\frac{s}{\sqrt{n}}} \\
&= \frac{\overline{x} - \mu}{s} \cdot \sqrt{n}
\end{aligned}
$$

We now define the index ($\hat{d}_i$), where $H_0 \colon \mu_i = M$ under Figure 3:

$$\hat{d}_i = \frac{\overline{x}_i - \mu_i}{s_i}$$

Observe that $\hat{d}_i$ is simply an unscaled $t$-statistic by $\sqrt{n}$:

$$\hat{d}_i = \frac{t}{\sqrt{n}}$$

Finally, we calculate the moments of $\hat{d}_i$. For the first moment $E\left[\hat{d}_i\right]$, since we know that $E[t] = 0$:

$$
\begin{aligned}
E\left[\hat{d}_i\right] &= E\left[\frac{t}{\sqrt{n}}\right] \\
&= \frac{1}{\sqrt{n}} \cdot E[t] \\
&= \frac{1}{\sqrt{n}} \cdot 0 \\
E\left[\hat{d}_i\right] &= 0
\end{aligned}
$$

For the second moment $E\left[\hat{d}_i\right]$, since we know that $Var(t) = \frac{\nu}{\nu-2}$:

$$
\begin{aligned}
Var\left(\hat{d}_i\right) &= Var\left[t \cdot \frac{1}{\sqrt{n}}\right] \\
&= \left(\frac{1}{\sqrt{n}}\right)^2 \cdot Var[t] \\
&= \frac{1}{n}\left(\frac{\nu}{\nu - 2}\right) \\
&= \frac{1}{n}\left(\frac{n-1}{n-3}\right) \qquad \nu = n - 1 \\
Var\left(\hat{d}_i\right) &= \frac{n-1}{n(n-3)}
\end{aligned}
$$

**2.2. Symbol and Interpretation**

The choice for the symbol $\widehat{d_i}$ is based on its statistical characteristics as a psychometric measure of item deviation and uncertainty. The "$d$" implies the measure as a specific derivation of Equation 9 while the hat "ˆ" accent signifies that it is an estimate of true psychometric deviation from the scale midpoint, in the same way as Cohen's $d$ is an estimate of the true effect size. The subscript $i$ denotes for the item $i$ and can be used to identify and distinguish the index of one item from another, yet a corrected $\hat{d}_i$ has the subscript $g$ because $\hat{d}_i$ is also multiplied by Hedges' correction making it mathematically analogous to Hedges' $g$.

The proposed index is an analytical derivation of Cohen's $d$ in the form of a standardized deviation index for a specific item $i$ that tells the extent of deviation among the respondent scores from the scale midpoint. Its theoretical center is found at 0, which represents the absence of deviation. Higher and lower values indicate that respondents tend to score higher and lower than usual within the psychometric scale, and whether its deviation is bad or good for the psychometric scale depends on the context from which it is interpreted.

**Figure 4**

*Visualization of Standardized Item Index (author-generated via ggplot2 R package)*

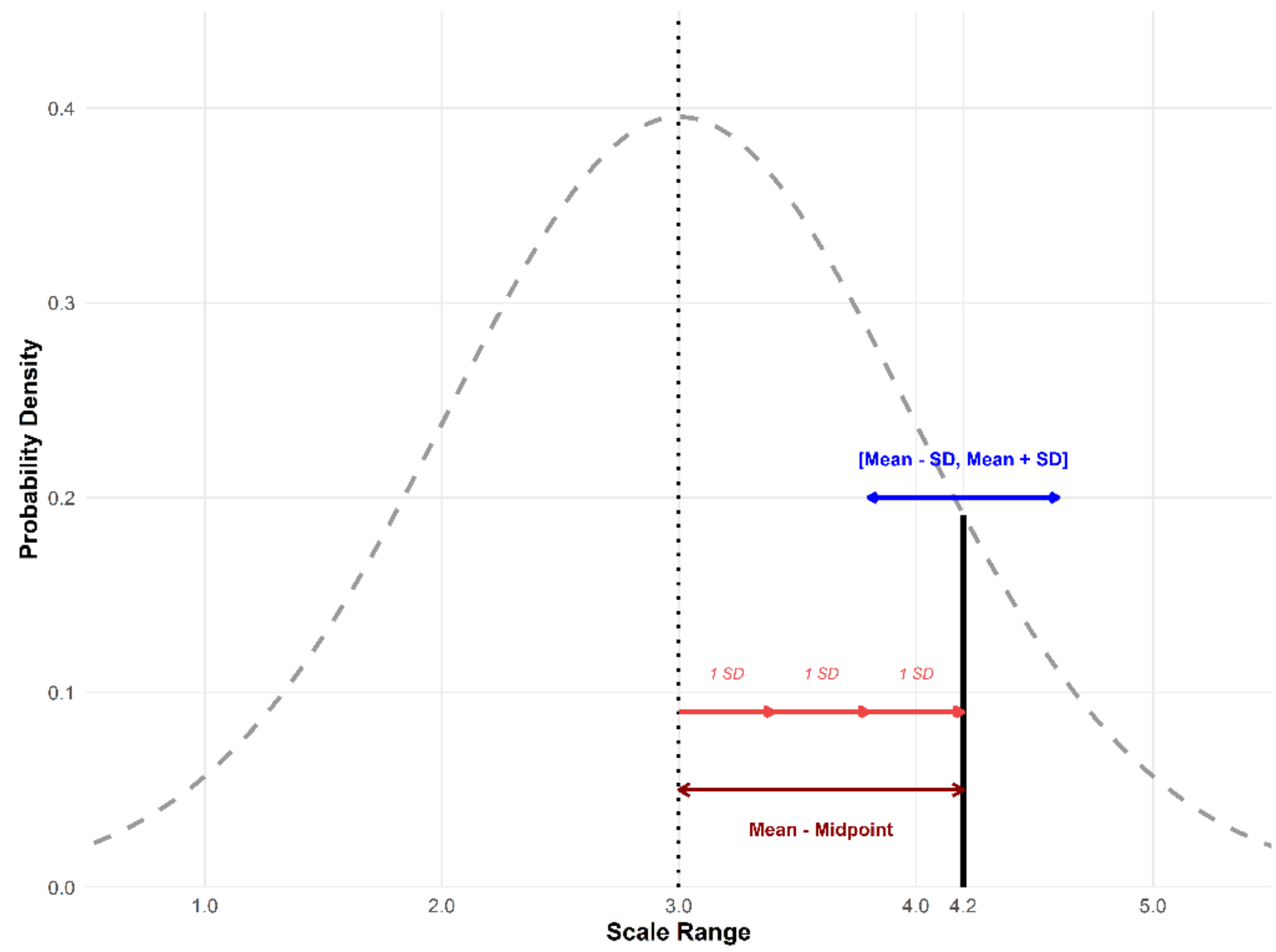

An intuitive way to understand what the index represents on the item is treating its value as telling the number of SD values needed to match the distance of $\overline{x}$ from $M$. Consider an example scenario of asking 30 respondents $n = 30$ to answer a particular item of a 5-point psychometric scale with $M = 3.0, \overline{x} = 4.2$, and $s = 0.4$. We can also observe the underlying theoretical distribution (Central $t$-distribution) of the index at $df = 29$.

Substituting values on the index gives us:

$$\hat{d}_i = \frac{4.2 - 3.0}{0.4} = 3$$

This means that it takes 3 SD units ($s = 3$) to match the theoretical distance on $\overline{x}$ from $M$. But what if the sign of $\hat{d}_i$ is negative? It means $\overline{x}$ is located on the left of the scale's midpoint, which validates a trait of $\check{d}$'s sign as accounting for the directionality of item deviation.

## 3. Results

### 3.1. On bounds of mean and standard deviation

We determine the numerator and denominator bounds of $\widehat{d_\iota}$ to allow us to understand its analytical behavior. For the denominator, Equation 7 can be applied since it provides analytical info involving variances. Given a standard deviation $s > 0$ and range $B - b$, we can then have:

$$s^2 \leq \frac{1}{4}(B-b)^2$$

Taking the square root of both sides yield

$$s \leq \frac{1}{2} \cdot |B-b|$$

.

Since $B - b > 0$,

$$s \leq \frac{B-b}{2}$$

Hence,

$$s \in \left(0, \frac{B-b}{2}\right] \tag{13}$$

While for the numerator, we simply derive its bounds from that of $\overline{x}$ itself,

Since we know $\overline{x} \in [b, B]$,

$$\begin{aligned}
&b \leq \overline{x} \leq B \\
&\Rightarrow b - \left(\frac{B+b}{2}\right) \leq \overline{x} - M \leq B - \left(\frac{B+b}{2}\right) \quad \text{Subtract by } M = \frac{B+b}{2} \\
&\Rightarrow \frac{b-B}{2} \leq \overline{x} - M \leq \frac{B-b}{2} \\
&\Rightarrow -\left(\frac{B-b}{2}\right) \leq \overline{x} - M \leq \frac{B-b}{2}
\end{aligned}$$

Hence,

$$\overline{x} - M \in \left[-\left(\frac{B-b}{2}\right), \frac{B-b}{2}\right] \tag{14}$$

### 3.2. On bounds of index

For the bounds of $\widehat{d_i}$, we will evaluate its convergence as $s$ approaches on both ends of its bounds. While we determined the both numerator and denominator, we prefer $s$ as the limiting variable because it involves less variables, making the analytical computation less complex.

As $s$ approaches its lower bound:

$$\begin{aligned}\lim_{s\to 0^+} \hat{d}_i &= \lim_{s\to 0^+} \frac{\overline{x} - M}{s} \\ &= \frac{\overline{x} - M}{0}\end{aligned}$$

We see that the convergence of $\widehat{d_i}$ depends on the value of the $\overline{x} - M$. Its limit behavior is a textbook case that a vertical asymptote exists at $x = 0$.

- Case 1: If $\overline{x} > M, \overline{x} - M > 0$. The limit is therefore approaching $+\infty$. However, given the upper bound of $\overline{x} - M$, $\hat{d}_i$ will theoretically converge to $\frac{B-b}{2}$.
- Case 2: If $\overline{x} < M, \overline{x} - M < 0$. The limit is therefore approaching $-\infty$. However, given the lower bound of $\overline{x} - M$, $\hat{d}_i$ will theoretically converge to $-\frac{B-b}{2}$.
- Case 3: If $\overline{x} = M, \overline{x} - M = 0$. The limit is therefore approaching 0.

Meanwhile, as $s$ approaches its upper bound:

$$\begin{aligned}\lim_{s\to \frac{B-b}{2}^-} \hat{d}_i &= \lim_{s\to \frac{B-b}{2}^-} \frac{\overline{x} - M}{s} \\ &= \frac{\overline{x} - \left(\frac{B+b}{2}\right)}{\frac{B-b}{2}}\end{aligned}$$

$$= \frac{\frac{1}{2}\left(2\overline{x} - (B + b)\right)}{\frac{1}{2}(B - b)}$$
$$= \frac{2\overline{x} - (B + b)}{B - b}$$

This expression has a similar mathematical structure as Equation 8 where $b$ and $B$ are the minimum and maximum values, respectively. Our goal now is to re-express it in terms of the min-max normalization expression.

$$
\begin{aligned}
&= \frac{2\overline{x} - (B + b)}{B - b} \\
&= \frac{2\overline{x} - B - b}{B - b} \\
&= \frac{2\overline{x} - B + (-2b + b)}{B - b} \qquad -b = -2b + b \\
&= \frac{2\overline{x} - 2b - (B - b)}{B - b} \qquad \text{Group terms.} \\
&= \frac{2(\overline{x} - b)}{B - b} - 1 \\
\lim_{s \to \frac{B-b}{2}^{-}} \hat{d}_i &= 2(x_{norm}) - 1
\end{aligned}
$$

Given $x_{norm} \in [0,1]$, observe that:

$$
\begin{aligned}
\text{If } x_{norm} &= 0, 2(0) - 1 = -1 \\
\text{If } x_{norm} &= 1, 2(1) - 1 = 1
\end{aligned}
$$

Therefore, $\lim \hat{d}_i$ as $s$ approaches its upper bound is bounded at $[-1,1]$.

### 3.3. Scale invariance

We determine whether $\hat{d}_i$ varies depending on the scaling parameter $k$. Let $X$ be a set of scores and $Y$ be a set of transformed scores such that $Y = kX$, where $k > 0$. $k > 0$ because we want to determine whether simply extending $(k > 1)$ or shrinking $(k < 1)$ the scale but a negative valued scale denotes little to no meaning (e.g. -5 point scale, -3 point scale, etc).

Given $\hat{d}_x$, we will first derive expressions for $\overline{y}$, $M_y$, and $s_y$:

$$\overline{y} = \frac{\sum_{i=1}^{n} k\, x_i}{n} = k\left(\frac{\sum_{i=1}^{n} x_i}{n}\right) = k\overline{x}$$

$$M_y = kM_x$$

$$\begin{aligned} s_y &= \sqrt{\frac{\sum_{i=1}^{n}(y_i - \overline{y})^2}{n-1}} \\ &= \sqrt{\frac{\sum_{i=1}^{n}(kx_i - k\overline{x})^2}{n-1}} \\ &= \sqrt{\frac{k^2 \cdot \sum_{i=1}^{n}(x_i - \overline{x})^2}{n-1}} \\ &= ks_x \end{aligned}$$

Hence,

$$\begin{aligned} \hat{d}_y &= \frac{k\overline{x} - kM_x}{ks_x} \\ &= \frac{k(\overline{x} - M_x)}{k(s_x)} \\ &= \frac{\overline{x} - M_x}{s_x} \\ &= \hat{d}_x \end{aligned}$$

### 3.4. Bias adjustment due to sample size

Since $\widehat{d_\iota}$ is an alternative case of Cohen's $d$ for one-sample t-test, Equation 10 also applies since Cohen's $d$ is known to be a biased estimator (see Equation 1) that tends to overestimate the true effect at small sample sizes. Hence, the correction $J$ at $\nu = n - 1$ is given by

$$\begin{aligned} J(n-1) &\approx 1 - \frac{3}{4(n-1)-1} \\ &\approx 1 - \frac{3}{4n-5} \end{aligned}$$

Hence, the corrected $\hat{d}_i$ (or $\hat{d}_g$) is simply multiplied by Hedges's correction:

$$\hat{d}_g = \check{R}\left(1 - \frac{3}{4n - 5}\right) \tag{15}$$

We now further explore the behavior of $\hat{d}_g$ as the sample size changes within the domain $n \in [2, \infty)$. $n = 1$ is excluded because $s$ cannot be calculated from a single observation. By Equation 12, we let $n$ be a continuous proxy to investigate the limit as it approaches either bound within the domain.

For the lower bound, we are interested in how $\hat{d}_g$ behaves as our sample size decreases.

$$\begin{aligned}\lim_{n \to 2^+} \hat{d}_g(n) \quad &= \check{R}\left(1 - \frac{3}{4n - 5}\right)\\ &= \check{R}\left(1 - \frac{3}{4(2) - 5}\right)\\ &= \check{R}\ (1 - 1)\end{aligned}$$

Hence,

$$\lim_{n \to 2^+} = 0 \tag{16}$$

This means $\hat{d}_i$ converges to 0 regardless of the value of $\hat{d}_i$.

Meanwhile, for the upper bound, we are interested in how $\hat{d}_g$ behaves as our sample size increases indefinitely.

$$\begin{aligned}\lim_{n \to \infty} \hat{d}_g(n) \quad &= \hat{d}_i\left(1 - \frac{3}{4n - 5}\right)\\ &= \hat{d}_i\left(1 - \frac{3}{\infty}\right)\\ &= \hat{d}_i\ (1 - 0)\end{aligned}$$

Hence,

$$\lim_{n \to \infty} = \hat{d}_i \tag{17}$$

This means that for large sample sizes, $\hat{d}_g$ converges to $\hat{d}_i$.

# 4. Discussion

## 4.1. The sampling distribution of index converges to a normal distribution.

While normal distribution is a fundamental concept in statistical theory, it is a misconceived and conflated term when dealing what needs to be normally distributed and why. It is important to emphasize here that we are interested in the sampling distribution being normally distributed, not the raw scores themselves, used to compute $\hat{d}_i$ (Midi et al., 2007; Kabir, 2013, Knief & Forstmeier, 2018) . The sampling distribution "converging" to a normal distribution at larger sample sizes as observed is supported by Equation 5 to establishes whether or not we are statistically "safe" to assume normality of our sampling distribution or not.

There are two suggested approaches to perform this check. The first method is a proxy approach: extract residuals after performing a one-sample $t$-test on the given item scores with the scale midpoint value as the true (population) value. You can then perform a QQ-plot check on the residual data to decide whether the sampling distribution converges or not. The second method is a direct approach: use the bootstrapping method (Mokhtar et al.,2023; Hair et al., 2019) to generate a sampling distribution of a set of bootstrap $\hat{d}_i$ values. You also set the number of bootstrap samples to generate and then count the number of bootstrap values that fall outside the acceptance region. However, establishing empirical thresholds and methodology from these approaches will be the scope of further research.

## 4.2. Standard deviation is a proxy of entropy.

In information theory, the entropy of a source measures its amount of uncertainty produced, which translates to the information available to be interpreted (Shannon, 1948; Rioul, 2021). Lower entropy values mean less noise (uncertainty) and therefore less informative, while higher entropy values indicate more uncertainty because it is less predictable and provides more information about the source.

Equation 11 shows that $(H)$ has a direct relationship with $\sigma$, and so the concept of uncertainty can be expanded to $\hat{d}_i$. Standard deviation measures the amount of uncertainty or noise from the given scores to provide information on how diverse the participants' responses are on a particular item. By accounting the diversity of their respondents, $\hat{d}_i$ allows for a more comparable estimate of deviation within other scale items and thus conceptualizes $\overline{x} - M$ as the "signal" of a particular measure of interest as a consequence. Beyond psychometrics, the conceptual link of entropy as a measure of variability also provides potential lead on establishing further applications of $\hat{d}_i$ on fields involving interactions between the signal and noise of a certain system.

**4.3. Displacement and standard deviation are bounded.**

From Equation 13, $\frac{B-b}{2}$ represents half of the actual scale length, and so the value of a standard deviation $(s)$ is restricted to never exceed this maximum bound. For instance, suppose we have a 5-point scale, where 1 is the minimum value, then an item $s$ on that same scale will never exceed 2 (because scale length is $5 - 1 = 4$). This also holds true regardless of the distribution of the raw scores.

The same goes for Equation 14, where the difference between the item mean and midpoint is restricted to be between two half-scale lengths. Given the symmetry on $\hat{d}_i$ values on Figure 4, the bounds also follow that same symmetrical property. By ensuring the bounds of the said inputs are finite, it simplifies further efforts on establishing empirical thresholds for the index across scales.

**4.4. The boundedness of index is dependent on standard deviation.**

The index, as well as its bounds, is functionally dependent on the standard deviation in the same way as entropy (Shannon, 1948), yet the nature of relationship is logically reversed. Lower standard deviation or entropy means less diversity (more consensus) among the responses (scores) of the respondents, and thus $\hat{d}_i$ becomes more accurate in directly measuring the raw deviation of the item mean to the midpoint as observed by its dependence on the value of the numerator. This validates the analytical ability of $\hat{d}_i$ as a diagnostic index of item deviation: Higher and lower $\hat{d}_i$ values indicate respondents are more likely to report higher and lower response than usual, respectively, which tells a general description on the participants' approach towards the item. On the contrary, when the standard deviation reaches its theoretical maximum, the domain of $\hat{d}_i$ becomes more conservative towards between $[-1,1]$. This shows that as participant variability affects the integrity of $\check{d}_i$, it still allows boundedness on where the values would lie.

This property ensures the robustness of $\check{d}_i$ across changes on the measured standard deviation, and also provides crucial consideration for the empirical thresholds since they are affected by the scale bounds and the actual standard deviation measured.

**4.5. The index is a unitless measure.**

$\hat{d}_i$ allows to be compared on the indices of other items without accounting for the actual scale's unit. As long as the items fall under a common scale (i.e. all item have common scale length with same scale response options), the interpretation for the index remains consistent. In other words, it doesn't matter whether we have 3-point or a 5-point scale, as well as having different response options (e.g. 0,1,2 or 1,2,3), the principles for interpreting the extent of deviation from $\hat{d}_i$ still holds.

In future efforts of establishing rigorous protocols for the index, this property directly validates that a universal interpretation on proposed empirical thresholds in future research is plausible.

**4.6. The index is descriptive but its corrected form is inferential.**

Literature widely acknowledges Hedges's $g$ as an unbiased alternative to Cohen's d for small sample sizes in providing a more accurate description of its true effect size (Zakzanis, 2001; Goulet-Pelletier & Cousineau, 2018; Becker, 2001; Delacre et al., 2021), which follows the same logic for $\hat{d}_i$ as holds true. However, Equation 16 reveals a nuanced property on Hedges's $g$ at varying sample sizes. Lower sample sizes increase the effect of correction on the original $\hat{d}_i$ up to a point where the value of $\hat{d}_g$ is simply 0, regardless of the value of $\hat{d}_i$. If we view $\hat{d}_g$ as a descriptive measure, the logic does not hold. How could having 2 observations guarantee your measure to detect zero bias, regardless of what the $\hat{d}_i$ really is? Apparently, $\hat{d}_g$ might not be intended to be a descriptive measure, but rather a measure with an inferential nature in the sense that it aims to establish whether the deviation we do observe is indeed practically significant, similar to how we would interpret effect sizes. While having 2 observations may mean that $\hat{d}_i$ measures the extent of standardized item deviation, $\hat{d}_g$ establishes whether your sample size is indeed sufficient to establish a claim of significant deviation. Equation 17 on the other hand shows that the correction reduces where $\hat{d}_g$ simply converges to $\hat{d}_i$. This confirms that the raw deviation that $\hat{d}_i$ is measuring is indeed a reliable estimate of the item's true deviation.

Common application of effect size assumes that null hypothesis is non-zero, following a non-central distribution. In other words, we are indeed assuming at the first place that it has an effect, but we are yet to validate its practical significance. This is in contrary on the application of Cohen's $d$ in this article implied by Figure 3. To the best of the author's reading, however, no existing literature has directly addressed the implicit property of Cohen's $d$ assuming a non-central null distribution. Hence, this property challenges our common understanding on the relation between Cohen's $d$ and Hedges's $g$ as origins of the derived index. Hedges's $g$ is not simply a "better", unbiased version of Cohen's $d$. In fact, if we view these two effect size measures from either side of hypothesis testing, we are clarified on why Cohen's $d$ is even seen as a biased estimate at the first place. In establishing our null hypothesis, the common idea is to set our true value to be 0, meaning we assume at first that there is no deviation and we just want to "confirm" it. This makes Cohen's d a sufficient and unbiased estimate within this context alone. However, establishing a null hypothesis does not need to always have a zero true value. We can assume that the null hypothesis is not zero, meaning there is deviation/effect and we also want to "confirm" it. This is where the Cohen's d "falls short" of overestimating the true effect size and therefore, Hedges' g becomes a necessary alternative (Hedges, 1981). The analogies from Cohen's $d$ and Hedges's $g$ also apply on $\check{d}_i$ and $\check{d}_g$, respectively.

For future protocols, $\hat{d}_g$ cannot replace $\hat{d}_i$ entirely, but rather it supplements the provided raw info of $d_i$ by answering one concern: Can we trust the info that $\hat{d}_i$ says given the sample size we have? If so, how close enough should $\hat{d}_g$ be to confidently trust $\hat{d}_i$?

## 5. Conclusion and Recommendations

The index is limited to numerical variables where arithmetic operators are doable, and mean and standard deviation establishes a precise meaning. Categorical variables (ordinal, nominal), on the other hand, do not have a precise, fixed distance across categories that causes variance-based measures to be unreliable and unstable. Hence, a categorical counterpart on the index is plausible and necessary.

It is also limited to Likert Scales since they are common survey scales being used in quantitative researchers, yet further efforts are needed to account particular properties that differentiate the said scales from its counterparts such as Guttman and Thurstone. While Likert Scales are theoretically ordinal, current pragmatic efforts are proposed to establish the convergence of psychometric properties of its scores towards continuous ones.